\documentclass[12pt]{article}
\usepackage{amsmath}
\usepackage{amssymb}
\usepackage{cite}

\usepackage{epsfig}

\psfull

\begin{document} 
\begin{titlepage}
\begin{flushright}
{
}
\end{flushright}              
\vspace*{\fill}
\begin{center}
{\Large 
\textsf{\textbf{Problems in resumming interjet energy flows with $k_t$ clustering}}}
\end{center}
\par \vskip 5mm
\begin{center}

  {\large \textsf{  A.~Banfi}} \\
  Cavendish Laboratory, University of Cambridge\\
  Madingley Road, Cambridge, CB3 0HE, U.K. \\
  {\large \textsf{ M.~Dasgupta}}\\
  School of Physics and Astronomy, University of Manchester \\
  Manchester M13 9PL, U.K.
\end{center}
\par \vskip 2mm
\begin{center} {\large \textsf{\textbf{Abstract}}} \end{center}
\begin{quote}
We consider the energy flow into gaps between hard jets. 
It was previously believed that the 
accuracy of resummed predictions for such observables can be improved by employing the 
$k_t$ clustering procedure to define the gap energy in terms of a sum of 
energies of soft jets (rather than individual hadrons) in the gap. 
This significantly reduces the sensitivity to correlated soft large-angle 
radiation (non-global leading logs), 
numerically calculable only in the large $N_c$ limit. While this is the case, 
as we demonstrate here, the use of $k_t$ clustering spoils the straightforward 
single-gluon Sudakov exponentiation that multiplies the non-global resummation. 
We carry out an ${\mathcal{O}}(\alpha_s^2)$ 
calculation of the leading single-logarithmic terms and 
identify the piece that is omitted by straightforward exponentiation. We compare our results with the full ${\mathcal{O}} (\alpha_s^2)$ result from the program EVENT2 to confirm our conclusions. For $e^{+}e^{-} \to 2$ jets and DIS (1+1) jets one can numerically resum these additional contributions as we show, but for dijet photoproduction and hadron-hadron processes further studies are needed.

\end{quote}
\vspace*{\fill}

\end{titlepage}


\section{Introduction}
Energy flow into gaps between hard jets  is a valuable source of information on many aspects of QCD. Since this radiation is typically soft, 
a perturbative calculation of the corresponding distribution contains large logarithms that need resummation. Comparisons of resummed perturbative 
estimates with data 
then enable us to test and further our knowledge of soft QCD dynamics.

Moreover the 
hadronisation corrections are large, making these spectra a useful testing 
ground for theoretical ideas about power corrections within say 
a dispersive model 
for the QCD coupling \cite{DMW}. Additionally at hadron 
colliders the activity away from 
jets has traditionally been used to study the soft underlying event and 
to refine models thereof \cite{Huston}, which will be an important component 
of physics at the LHC.

In the present letter we analyse the first aspect alone, that of 
perturbative resummation. This resummation is challenging since the 
leading single-logarithms (there are no collinear enhancements due to 
the fact that we are away from the hard jets) are generated both by direct angular-ordered emission into the gap by the hard parton (jet) system as 
well as multiple energy-ordered correlated emission by a complex ensemble of soft emissions outside the gap \cite{Dassal1,Dassal2}, in addition to the hard jets. 
This latter piece, which cannot yet be computed analytically and 
more worryingly beyond a large $N_c$ approximation, 
is known as the non-global component of the result. The non-global term  causes a much 
stronger suppression of the 
soft energy flows than that obtained by vetoing direct emission off hard partons, into the gap, the Sudakov or Bremsstrahlung component of the answer \cite{Dassal2}. 
The final result, for the cross-section with gap energy less than $Q_\Omega$  
can be expressed as \cite{Dassal1,Dassal2,BMS}
\begin{equation}
\label{result}
\Sigma(Q,Q_{\Omega}) = \exp\left[-R(Q/Q_{\Omega})\right] S\left (Q/Q_{\Omega} \right ).
\end{equation}.

In the above $\exp[-R]$ is the Sudakov term obtained by exponentiating the single gluon contribution and accounting for gluon branching to reconstruct the scale of the 
running coupling, while $S$ is the non-global part of the answer obtained 
by 
running a large $N_c$ Monte-Carlo program that encodes soft evolution of 
a system of dipoles to single-log (SL) accuracy, for emission into the gap $\Omega$. 
Equivalently 
for such away-from--jet energy flows one has to solve numerically a non-linear evolution equation \cite{BMS} obtained in the large $N_c$ limit.

Given that only large $N_c$ approximations of the non-global component are 
calculable at present in conjunction with the fact that it dominates the full result at smaller $Q_{\Omega}$ values, it is important to reduce the sensitivity of the measurement to this effect. One method of doing so is to study event-shape-energy--flow correlations \cite{BKS,DokMar} 
where measuring an event shape $V$ 
outside the gap at the same time as the energy flow $Q_{\Omega}$ inside it leads to non-global logarithms in the ratio $V/Q_{\Omega}$. Thus choosing $V \sim 
Q_\Omega$ the magnitude of the non-global effects is reduced. However this procedure is more complex to implement in the case of several hard partons (e.g jet production in hadron-hadron collisions) and is quite restrictive experimentally, amounting essentially to studying a different observable. 

Another version of the measurement was suggested in Refs.~
\cite{AS1,AS2}, to reduce the impact of non-global logarithms. There it was 
shown that defining the energy $Q_{\Omega}$ as the sum of energies of soft (mini) {\it{jets}} inside the gap region significantly reduced 
the non-global component of the result. 
Defining the jets via a $k_t$ clustering procedure \cite{ktpaper}, which is also common practice experimentally \cite{H1,ZEUS}, had the effect of pulling soft emissions out of 
the gap region by clustering with 
harder emissions outside. 
As was shown in the simple case of $e^{+}e^{-} \to 2 $ jets \cite{AS1}, it is still possible for emissions near the centre of the gap to escape clustering. However since these emissions are well separated in 
rapidity from their nearest neighbours outside the gap (to escape clustering) the magnitude of non-global effects is reduced. 
This is because the bulk of the non-global piece arises in the region 
when the emitted 
soft gluon does not have too large an opening angle relative to those involved in the  emitting ensemble \cite{Dassal2}. Forcing a relatively 
large opening-angle/rapidity separation between the softest gluon and the harder emitters, as is required to escape clustering, does reduce the size of the non global effects \cite{AS1}.

We would like to point out, however, that using $k_t$ clustering impacts 
the general form Eq.~\eqref{result} and does not leave the primary emission Sudakov piece 
 $\exp[-R]$ unchanged as has been assumed till now \cite{AS1,AS2}. In fact we find the exact calculation of $R$ becomes non-trivial at higher orders since it depends at $n^\mathrm{{th}}$ order on the $n$ gluon geometry and the use of the clustering algorithm. 
The departure of $R$ from its naive one-gluon form starts with two gluons 
and the resulting piece does not have the properties of non-global logs , neither in the colour structure nor the dynamical properties. This conclusion is unfortunate especially in the case of dijet photoproduction \cite{AS2} or hadron-hadron studies where the missing piece we compute at leading order here, will have a highly non-trivial colour structure since it reflects the colour flow of the primary emission piece, computed e.g in \cite{AS2,BKS2}. This would impact accurate theoretical studies of such observables even though the non-global part is reduced.

The current letter is organised as follows. In the subsequent section we identify the problem with naive one-gluon exponentiation to obtain the supposed 
Sudakov part of the answer, with $k_t$ clustering. We compute the piece that will 
be missed by one-gluon exponentiation, at leading order ${\mathcal{O}}(\alpha_s^2)$. We then provide the full answer up-to order $\alpha_s^2$ , for the leading logarithms, and show via comparisons with EVENT2 \cite{Nason} that the extra piece we compute is needed to agree with fixed order estimates while 
the expansion to 
order $\alpha_s^2$ of the Sudakov term, as currently computed in the literature, is insufficient. We conclude by pointing out that while analytical control over the Sudakov term is lost, due to clustering, it is possible in the simpler cases $e^{+}e^{-} \to$ two jets and DIS (1+1) jets, to numerically compute 
the additional 
piece at all orders, with existing programs \cite{AS1}, as we show. For higher jet topologies such as those studied in Ref.~\cite{AS2} it may only be possible to numerically compute this term in the large $N_c$ limit, thereby reducing the accuracy of the theoretical results compared to current expectations. 

\section{Independent soft gluon emission}
We specialise to the process $e^{+}e^{-} \to 2$  jets 
purely for reasons of simplicity. 
We perform our calculation for emission into a rapidity slice of width 
$\Delta \eta$, centred at  zero rapidity with respect to the jet (thrust) axis, 
although similar considerations apply to any geometry one may choose for the inter-jet region $\Omega$.  
We first consider the Sudakov exponent generated by assuming exponentiation of single-gluon emission into the gap $\Omega$ (c.f. Eq.~(3.9) of Ref.~\cite{AS1}):
\begin{equation}
\label{primary}
\Sigma_{\Omega,P}(t) =\frac{1}{\sigma} \int_0^{Q_\Omega} \frac{d\sigma}{d \omega} d\omega  = e^{-4 C_F t \Delta \eta},
\end{equation}
with $t= \int_{Q_\Omega}^{Q/2} \frac{dk_t}{k_t} \alpha_s(k_t) =
\frac{1}{4 \pi \beta_0} \ln \frac{1}{1-2 \lambda}$ , where we used the one-loop running coupling to obtain up-to SL accuracy and defined $\lambda = \beta_0 \alpha_s(Q) \ln \frac{Q}{2 Q_\Omega}$. 
 
We now argue that the above form which exponentiates the single gluon (leading-order) term 
is not correct starting from two gluon level. \footnote{By this we mean that even after accounting for non-global logarithms, the exponentiation of the single gluon result 
still does not capture completely the remaining leading logarithms.} 
Consider two soft emissions $k_1$ and $k_2$ that are ordered in energy $\omega_1 \gg \omega_2$, with $\omega_1$ outside the gap and $\omega_2$ inside. The situation is reminiscent of the correlated or non-global configuration at leading order. 
However that part 
pertains to the $C_F C_A$ correlated gluon emission 
piece of the matrix element that is free from collinear singularities. 
In the present case we shall examine instead 
the {\it{independent emission}} $C_F^2$ part of the two gluon matrix element along with the corresponding virtual corrections. We have for the 
independent  emission of two real gluons by a dipole $ab$ \cite{DMO}:

\begin{equation}
\label{real}
M^2(k_1,k_2) = C_F^2 W_{ab}(k_1)W_{ab}(k_2) =4 C_F^2 \frac{(ab)}{(ak_1)(bk_1)} \frac{(ab)}{(ak_2)(bk_2)},
\end{equation}
where $W_{ab}(k_1)$ represents the emission of $k_1$ off the hard dipole $ab$ 
and similarly for $k_2$.

Now we examine the region where the two real soft 
gluons $k_1$ and $k_2$ are clustered 
by the jet algorithm. This happens when 
\begin{equation}
\label{distance}
(\eta_1-\eta_2)^2+(\phi_1-\phi_2)^2 <  R^2,
\end{equation}
where $\eta$ and $\phi$ denote as 
usual rapidity and azimuth of the partons measured with respect to the axis defined by the back-to--back partons $a$ and $b$ and $R$ is the radius parameter, usually set equal to one in experiment \cite{H1,ZEUS}. 

Since $k_1$ is outside the gap , $k_2$ is clustered into $k_1$ and pulled out of the gap. This configuration then does not contribute to the gap energy distribution $d\sigma/d\omega$ (see Eq.~\eqref{primary}). However now let us take the situation where we have $k_1$ virtual and $k_2$ as a real emission. Then $k_2$ is not clustered away by the algorithm and this configuration contributes with weight \cite{DMO}:
\begin{equation}
\label{virt}
M^2(k_{1,{\mathrm{virtual}}},k_2) = -  C_F^2 W_{ab}(k_1)W_{ab}(k_2), \, \omega_1 \gg \omega_2.
\end{equation}
We thus have complete cancellation between the purely real and real-virtual 
terms, Eqs.~\eqref{real} and \eqref{virt} in the region where $k_2$ is not removed by clustering. However in the angular region mapped by Eq.~\eqref{distance} {\it{only}}
the one-real-one--virtual term will contribute, since $k_2$ is in the gap.

This contribution can then be computed as below. The four-vectors involved are 
\begin{eqnarray}
a &=& \frac{Q}{2} (1,0,0,-1) \\
b &=& \frac{Q}{2}(1,0,0,1)\\
k_1 &=& k_{t1}(\cosh\eta_1,0,1,\sinh\eta_1)\\
k_2 &=& k_{t2}(\cosh\eta_2,\sin(\phi),\cos(\phi),\sinh\eta_2)\\
\end{eqnarray}
where we have exploited the freedom to set $\phi_1=0$ and $Q$ is the $e^{+}e^{-}$ centre-of--mass energy. We have also neglected the recoil of the hard partons $a$ and $b$, against the soft emissions $k_1$ and $k_2$, which is valid for 
our aim of extracting the leading logarithms.

Then the leftover real-virtual two-loop contribution reads 
(we compute the coefficient of $\left (\alpha_s/2\pi \right)^2$)
\begin{equation}
\label{veto}
C_2^{{\mathrm{primary}}}= 16 C_F^2 \int\frac{dx_2}{x_2} \int \frac{dx_1}{x_1} \Theta\left(x_2-\frac{2 Q_\Omega}{Q} \right) \int_{k_1 \notin \Omega}d\eta_1\int_{k_2 \in \Omega} d\eta_2 \frac{d\phi}{2 \pi} \Theta(R^2-(\eta_2-\eta_1)^2-\phi^2).
\end{equation}

The above equation requires some explanation. 
We have introduced the dimensionless scaled transverse momenta 
$x_{1,2} =2 k_{t1,t2}/Q$ and restricted the region such that virtual emission $k_1$ is integrated outside the gap region while real emission $k_2$ inside. We have also inserted a step function that ensures that we are integrating over the 
region of Eq.~\eqref{distance}, where the corresponding double real emissions would be clustered and 
$k_2$ would be pulled out of the gap. The additional step function involving 
$x_2$, that constrains the gap energy, is the usual one that corresponds to 
computing the cross-section for events with energy in the gap greater than $Q_{\Omega}$. This, by unitarity, is trivially 
related to that for events with gap energy 
less than $Q_{\Omega}$. From the latter quantity the distribution is directly obtained by differentiation with respect to $Q_\Omega$. 
We have denoted this term $C_2^{{\mathrm{primary}}}$ as it is a second order in $\alpha_s$ piece that has the colour structure and matrix element for independent emission from the primary dipole $ab$. However it is not derived by expanding the standard Sudakov result to order $\alpha_s^2$ and is a companion to the non-global correction term $S_2$ of \cite{AS1}, but with different functional 
properties and colour structure.

Performing the integration over $\phi$ in Eq.~\eqref{veto} we get
\begin{equation}
C_2^{{\mathrm{primary}}} = \frac{16}{\pi} C_F^2 L^2 \int_0^R du \, 
{\mathrm{min}}(u,\Delta \eta)\, \sqrt{R^2-u^2}.
\end{equation}

Choosing for instance values of the gap size $\Delta \eta \geq R$ we get 
\begin{equation}
\label{new}
C_2^{{\mathrm{primary}}} = \frac{16}{3 \pi} C_F^2 L^2 R^3.
\end{equation}
with $L= \ln \frac{Q}{Q_\Omega}$.
Alternatively choosing a smaller gap $\Delta \eta \leq R$ we get 
\begin{multline}
\label{new2}
C_2^{{\mathrm{primary}}} = \frac{16}{\pi} C_F^2 L^2 \left [\frac{1}{3}\left(R^3-(R^2-(\Delta \eta)^2)^{3/2}\right) +\frac{\Delta \eta}{2} \left (\frac{\pi R^2}{2} 
\right . \right. \\ 
\left . \left. -\Delta \eta\sqrt{R^2-(\Delta \eta)^2}-R^2 \tan^{-1} \left(\frac{\Delta \eta}{\sqrt{R^2-(\Delta \eta)^2}} \right) \right) \right ].
\end{multline}

It is clear that 
although this piece has the same colour structure as that for independent 
two gluon emission, it cannot arise from expanding the single-gluon generated Sudakov Eq.~\eqref{primary}. The expansion of the naive Sudakov would give a term independent of $R$ and which goes as $(\Delta \eta)^2$ at ${\mathcal{O}} 
\left ( \alpha_s^2 \right) $. 

In the following section we shall show that the expansion of the Sudakov 
Eq.~\eqref{primary} needs to be supplemented with the results 
Eqs.~\eqref{new},\eqref{new2} as appropriate, as well 
as the correlated non-global $C_F C_A \alpha_s^2 L^2$,  piece computed in 
\cite{AS1}, in order to agree with the full $\alpha_s^2 L^2$ result, for $\Sigma(Q/Q_\Omega)$ generated by the program EVENT2.

\section{Full ${\mathcal{O}} (\alpha_s^2)$ result and comparison to EVENT2}
First we expand the Sudakov exponent Eq.~\eqref{primary} to 
${\mathcal{O}}({\alpha_s^2})$. The result is 
\begin{equation}
\label{expan}
\Sigma_\Omega(Q,Q_\Omega) = 1- \bar{\alpha_s} L (4 C_F \Delta \eta) +
\bar{\alpha_s}^2 L^2 \left(8 \left (\Delta \eta\right)^2 C_F^2 -\frac{22}{3 } \Delta \eta C_FC_A+\frac{4 C_F n_f \Delta \eta}{3} \right ),
\end{equation}
where we denoted $\bar{\alpha_s} = \frac{\alpha_s}{2\pi}$

An additional $C_F C_A \alpha_s^2 L^2$ term is indeed the non-global term computed in Ref.~\cite{AS1}. 
We compute this piece numerically for different values of the parameters $\Delta \eta$ and $R$ and add it to the result from Eq.~\eqref{expan} for comparison with the fixed order program EVENT2.  For example, with $R=1$ and $\Delta \eta=1$ one gets 
$S_2 = -1.249 \,C_F C_A $ where $S_2$ is the first coefficient of the non-global 
log contribution $S=1+\sum_{n=2} S_n t^n$, with $t$ defined as before.

\begin{figure}

\epsfig{file=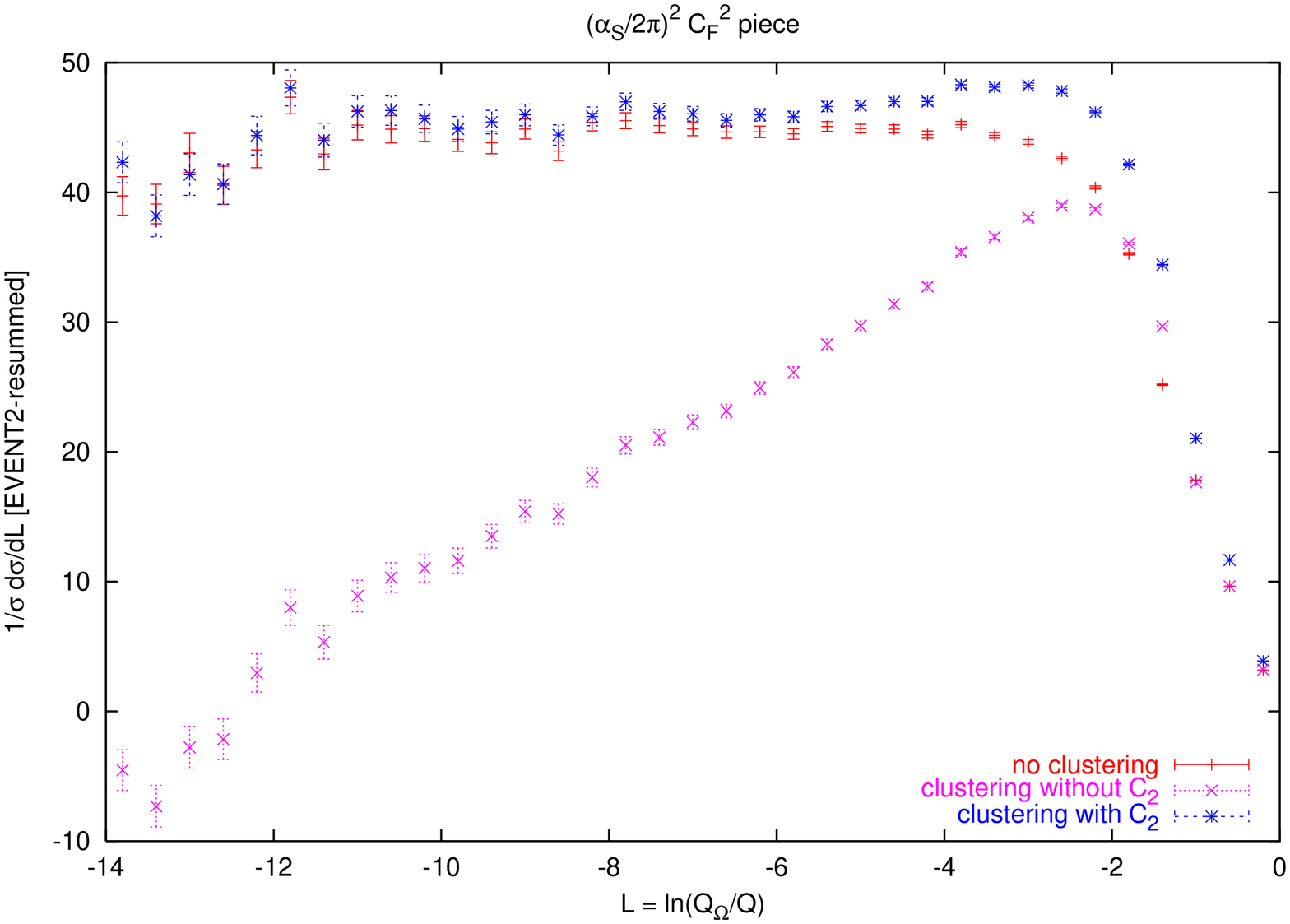, width =0.7\textwidth, angle=270}

\caption{Comparison of the $C_F^2 \alpha_s^2 L$ term produced by EVENT2 and the analytical calculation (referd to as resummed since it is derived by expanding 
the naive Sudakov 
resummation to NLO) with and without $C_2^{\mathrm{primary}}$. The figures are for $R=1$ and $\Delta \eta=1.0$ (above) and $\Delta \eta=0.5$ (below). The agreement for the unclustered case is also shown for comparison.}
\epsfig{file=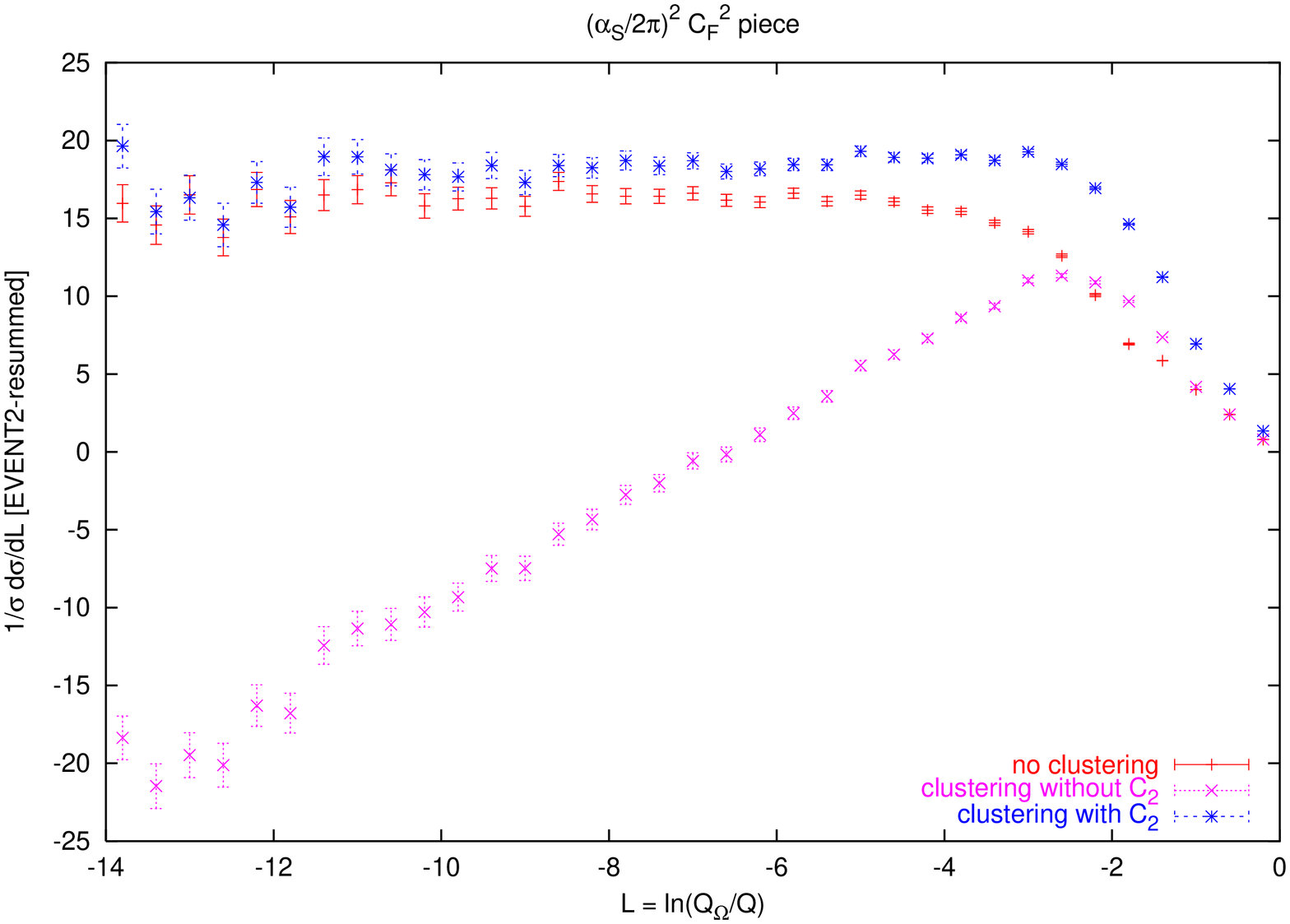, width =0.7\textwidth,angle=270}
\label{cf}
\end{figure}

\begin{figure}

\epsfig{file=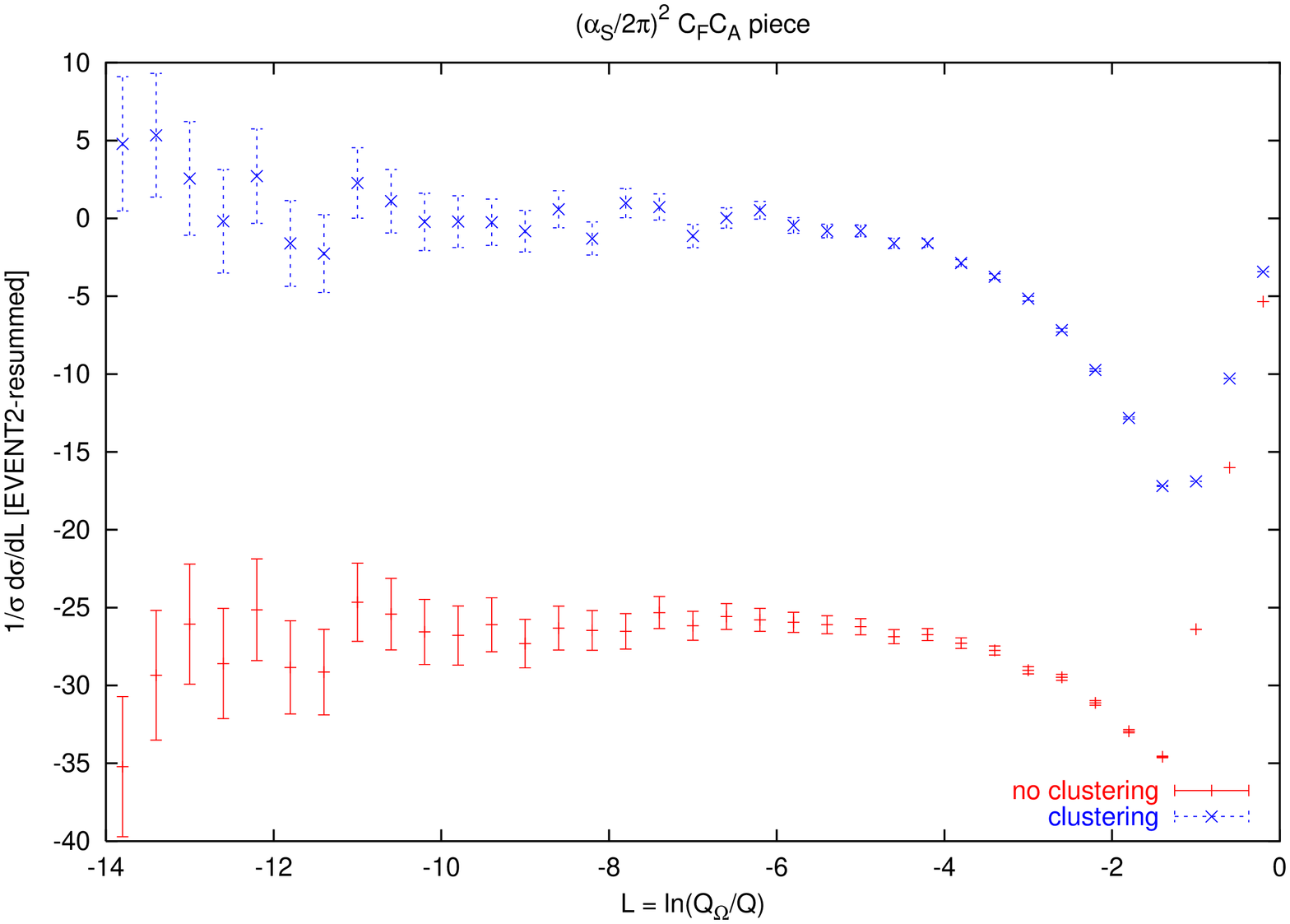, width =0.7\textwidth, angle=270}
\caption{Comparison of the $C_F C_A $ (above) and $C_Fn_f $ $ \alpha_s^2 L$ 
term (below) produced by EVENT2 and the expanded Sudakov result, 
supplemented with non-global logs for the $C_F C_A$ term. 
The figures are for $R=1$ and $\Delta \eta=1.0$ and 
and as we expect the difference between the exact and resummed result 
expanded to NLO is a constant at large $L$.}
\epsfig{file=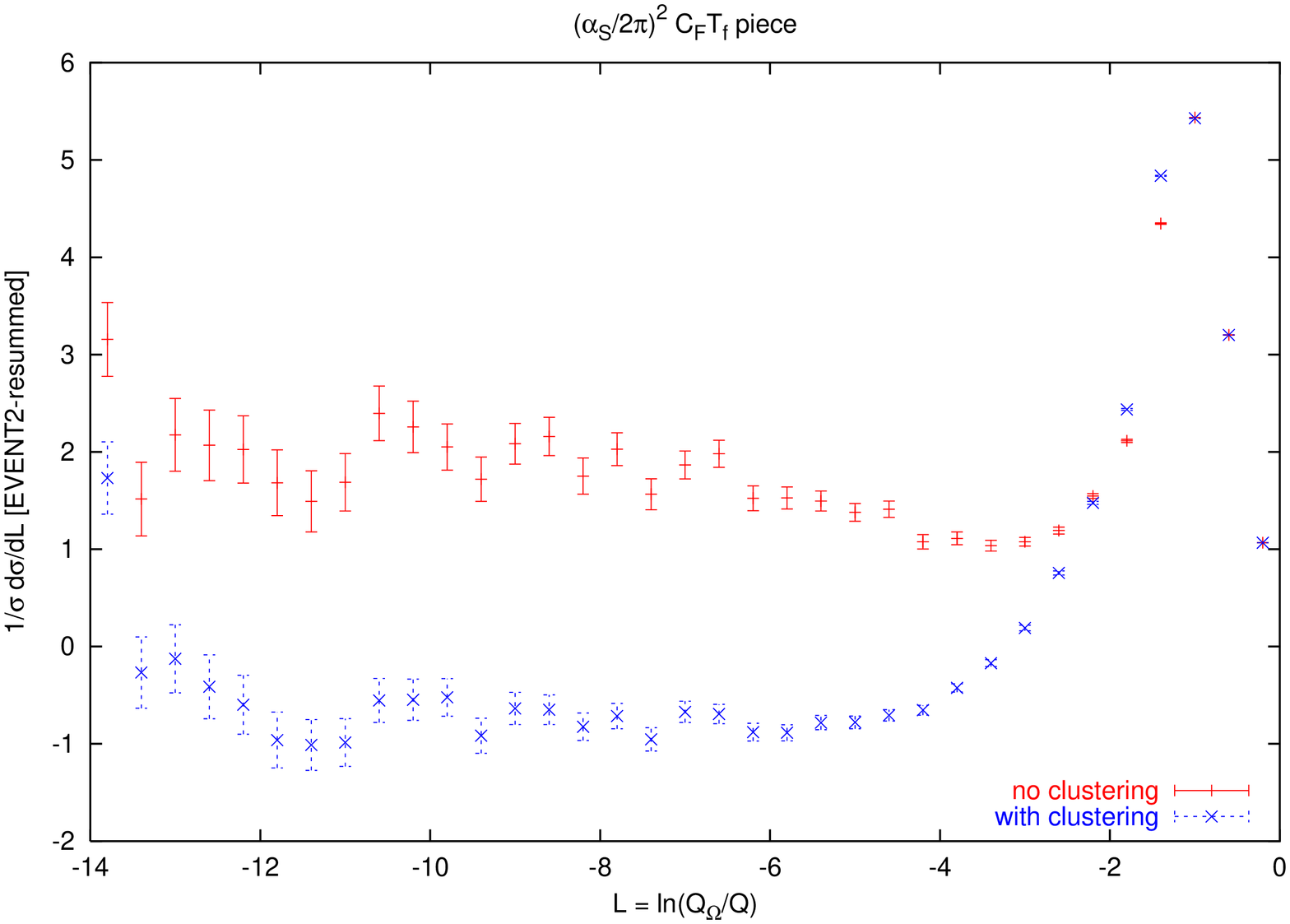, width =0.7\textwidth,angle=270}
\label{ca}
\end{figure}

The comparison to EVENT2 for the distribution $d\sigma/dL$ is shown in 
Fig.~\ref{cf} for the $C_F^2 \alpha_s^2 L$ term, with $L=\ln Q_\Omega/Q$ for 
$R=1$ and $\Delta \eta=1$, as well as $\Delta \eta=0.5$. If all leading (single) logarithms in the 
integrated quantity $\Sigma(Q,Q_\Omega)$ are correctly accounted for by the 
resummed result Eq.~\eqref{primary}, we would expect the difference between 
the EVENT2 results and the expansion to NLO of the resummation, to be a constant at small $Q_\Omega$ corresponding to large (negative) $L$. As we see this is only the case when $C_2^{\mathrm{primary}}$ is 
included by adding it to the expansion of the Sudakov Eq.~\eqref{primary}. We have considered different values of $R$ and $\Delta \eta$ as mentioned, for example, in the caption for Fig.~\ref{cf}. 
The comparison for other colour channels $C_F C_A$ and $C_F T_{R}n_f$ shows 
agreement with EVENT2 (see Fig.~\ref{ca}) which reflects the fact that only the $C_F^2$ channel, corresponding to independent emission, is 
incorrectly described by Eq.~\eqref{primary}.

\section{All orders and conclusions}
We conclude by pointing out that in the simple cases of 
$e^{+}e^{-} \to$ two jets and DIS (1+1) jets, the additional terms we describe here can be accounted for numerically, at all orders. This is done by using 
the Monte-Carlo program for large $N_c$ dipole evolution developed in 
\cite{Dassal1} and implemented with the $k_t$ clustering in \cite{AS1}.
By demanding multiple emissions from the primary dipole alone and restoring the 
colour factor for independent emission by changing $C_A/2 \to C_F$ one obtains the result for primary emissions only, in the full theory.  

\begin{figure}

\epsfig{file=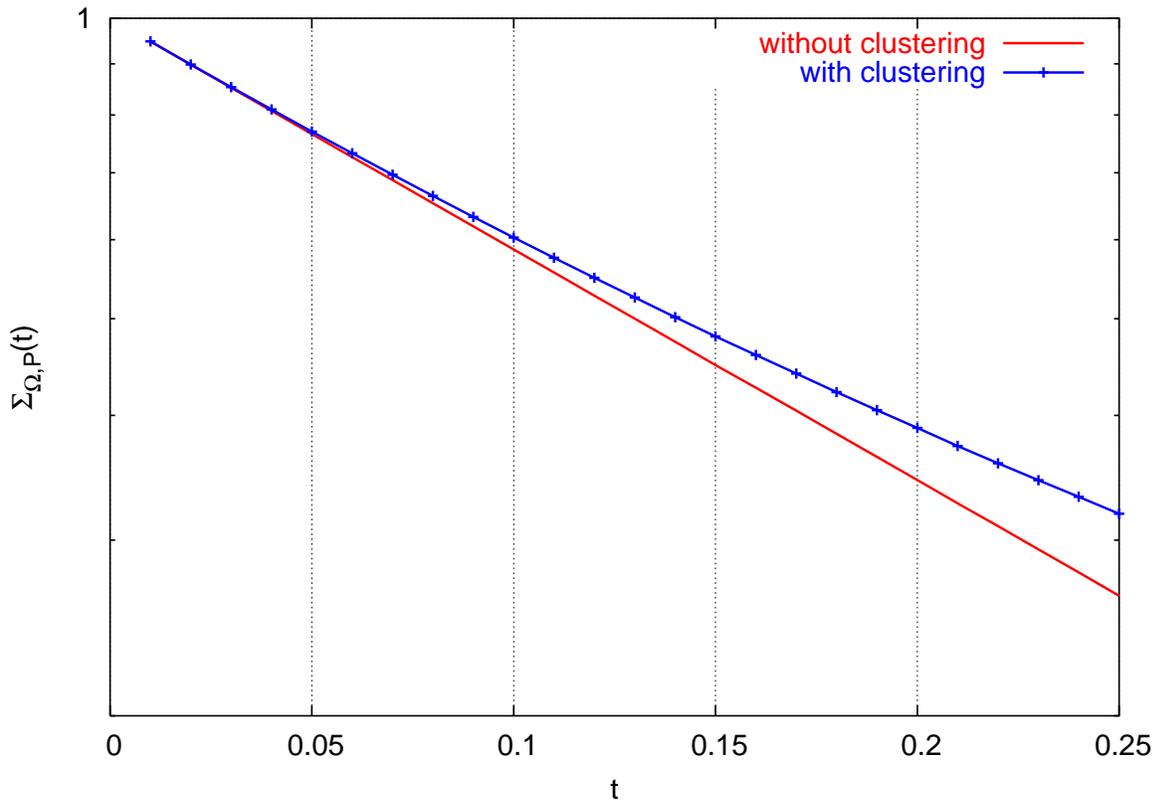, width =0.7 \textwidth, angle=270}

\caption{The results for the primary emission resummation with and without 
$k_t$ clustering for $R=1 \, , \Delta \eta=1$, using an adaptation of the program used for Ref.~\cite{AS1}. 
As can be seen, the clustering affects the primary emission term 
and the effect for $t=0.25$ is an increment of over 30 \%.}
\label{allord}
\end{figure}

Using this procedure we see that the primary emission result, 
with $k_t$ clustering, differs 
from the Sudakov result generated by single gluon exponentiation, which is valid in the unclustered case. 
In Fig.~\ref{allord}, we show the results we obtain for primary emission 
with clustering and the Sudakov (unclustered) result.
The discrepancy grows with the single-log 
evolution variable $t=\frac{1}{4 \pi \beta_0} \ln \frac{1}{1-2 \lambda}$ 
where $\lambda = \beta_0 \alpha_s(Q) \ln \frac{Q}{2Q_\Omega} $ and for $t=0.25$ we note an increase of around $30 \%$ on inclusion of the terms we 
describe, 
that start with $C_2^{\mathrm{primary}}$ computed analytically here. 

We wish to clarify that the full result in the large $N_c$ approximation, 
including the effect we point out here,  is readily 
obtained by the method described in Ref.~\cite{AS1} and in fact computed there. 
However its separation into primary and non-global components (and restoring the proper colour factors where possible) needs to be done with care, keeping in mind our findings. 

The procedure to generate the most accurate theoretical 
results for the 
$e^{+}e^{-} \to$ two jets and DIS (1+1) jets is the following. 
We take the results as generated by the code used for Ref.~\cite{AS1}, for a given gap geometry. This is the full result in the large $N_c$ approximation 
and we divide it by the result obtained using the same code for {\emph{primary emissions alone}} (rather than dividing by the naive Sudakov result), which takes as the only source for emissions the original hard dipole, e.g. the 
outgoing $q\bar{q}$ pair in $e^{+}e^{-}$ annihilation. The result of this division is the non-global piece in the large $N_c$ limit. We can then make use of the  fact that one can easily compute the exact ${\mathcal{O}}(\alpha_s^2)$ 
non-global term with proper $C_F C_A$ colour 
factor and parameterise the non-global Monte-Carlo result, as a function of $t$, in a form that retains the correct colour structure for the 
leading $\alpha_s^2 \ln^2 Q/Q_\Omega$ non-global term (see e.g. Ref.~\cite{Dassal1}). 
This is the non-global result, with the large $N_C$ approximation 
starting only from ${\mathcal{O}}\left(\alpha_s^3 \ln^3 Q/Q_\Omega \right)$ 
terms. 
The overall result is obtained by multiplying the resultant parameterised form 
with the primary emission result, as computed here with full colour factors. 
The large $N_c$ approximation is then confined to the non-global term 
and starting from the next-to--leading such piece ($S_3$ in the notation of \cite{AS1}). It is thus 
still an important finding 
that the non-global logarithms are reduced considerably by $k_t$ clustering as 
demonstrated in Ref.~\cite{AS1}, since this potentially reduces the 
impact of unknown non-global terms beyond the large $N_C$ approximation.
However the correct procedure for identifying the primary and non-global pieces, pointed out here, 
must be accounted for while comparing to experimental data to enable 
accurate phenomenological studies.

In the case of dijet photoproduction, studied e.g. in \cite{AS2}, and 
gaps between jets in hadron-hadron processes, it is less 
straightforward to account for the missing independent emission terms we point out. 
They will have a complex colour structure and existing large $N_c$ numerical 
programs cannot be employed to generate the full answer beyond the large $N_c$ limit. 
This would mean that the accuracy of the resummed result is limited not just by the unknown beyond-large--$N_c$ non-global logs but similarly 
in the primary emission terms which are not reduced by the use of clustering. 
In these cases further studies are therefore required to 
account correctly for the missing primary emission 
terms before one can argue that use of the 
clustering method mitigates the uncertainty involved in the theoretical predictions, by reducing the non-global component significantly. This is currently work in progress \cite{corbandas}.

\par \vskip 1ex

\noindent{\large\bf Acknowledgments}
We would like to thank Mike Seymour for useful discussions 
concerning this article and the work described in Ref.~\cite{AS1} and Rob Appleby for 
supplying us with the numerical code used in Ref.~\cite{AS1}. 
We also thank Gavin Salam for helpful comments.



\end{document}